\documentclass[aps,prl,twocolumn,superscriptaddress]{revtex4-2}

\usepackage{graphicx}	
\usepackage{siunitx}	
\usepackage{color}		
\usepackage{physics}	
\usepackage[version=4]{mhchem}
\def\({\left(}
\def\){\right)}

\def\be{\begin{eqnarray}}
\def\ee{\end{eqnarray}}

\begin{document}

\title{Effect of deuteration on methane cluster stability}
\author{{\AA}ke Andersson$^{\ddagger}$}
\affiliation{Department of Physics, University of Gothenburg, 41296 Gothenburg, Sweden}
\author{Klavs~Hansen$^*$}
\affiliation{Center for Joint Quantum Studies and Department of Physics, School of Science,
Tianjin University, 92 Weijin Road, Tianjin, 300072, China.}
\email{klavshansen@tju.edu.cn}
\homepage{www.klavshansen.cn}
\author{Siegfried~Kollotzek$^{\dagger}$}
\affiliation{Institut f\"ur Ionenphysik und Angewandte Physik, Universit\"at Innsbruck,
Technikerstr. 25, Innsbruck, 6020, Austria}
\author{Johannes~Reichegger}
\affiliation{Institut f\"ur Ionenphysik und Angewandte Physik, Universit\"at Innsbruck,
Technikerstr. 25, Innsbruck, 6020, Austria}
\author{Paul~Scheier}
\affiliation{Institut f\"ur Ionenphysik und Angewandte Physik, Universit\"at Innsbruck,
Technikerstr. 25, Innsbruck, 6020, Austria}
\date{\today}

\begin{abstract}
	The evaporation from clusters of mixed deuterated and protonated methane molecules 
	has been studied experimentally and theoretically. 
	Only loss of the two components \ce{CH4} and \ce{CD4} is observed, showing the absence of 
	hydrogen scrambling.
	The branching ratio of the two channels scales with the mole fractions of the 
	isotopologues, corresponding to a fixed ratio of 1.5 for the \ce{CH4} and \ce{CD4} 
	rates.
	The number is explained by the combined effect of the zero-point energies found 
	in quantum chemical calculations and the use of detailed balance to describe the 
	evaporation process, the latter by incorporating the moment of inertia of the two 
	isotopologue molecules and their rotational degrees of freedom.
	The difference in evaporative activation energies is determined to be 6.1~\% and 6.8~\% 
	from two experimental determinations and 8.7~\% theoretically with B3LYP+GD3BJ, 
	corresponding to 5--7~meV.
	The same theory finds the CH$_4$ activation energy to be 83.8 meV, close to the 
	experimental bulk value of 84.7 meV.
  Differences in Helmholtz free energies of decay products are extracted from the 
	experimental data, yielding the molecular mixing as the only contribution to the 
	entropy term.
\end{abstract}

\maketitle
{\footnotesize
	\noindent
\noindent
$^{\ddagger}$Present address: Department of Physics and Nanoscience Center, 
University of Jyv{\"a}skyl{\"a}, Finland
	
\noindent	
$^{\dagger}$Present address: Department of Radiation Oncology, Medical University 
of Innsbruck, Austria}

\section{\label{intro}Introduction}

Deuterium substitution is a widely used technique for probing reactions of biological 
molecules and of other hydrogen-containing molecules that exhibit hydrogen
bonding, because hydrogen scrambling is easily monitored with mass spectrometry.
The clearest form of hydrogen bonds is arguably found in water.
Collisional experiments on water clusters with heavy water molecules have shown a 
strong degree of hydrogen scrambling on time scales as short as tens of 
microseconds, at least for sufficiently large clusters~\cite{AndersssonPCCP2008,RydingPCCP2011}.
More recently, it was shown that water reaches bulk thermal properties already at 
cluster size $N=9$~\cite{HansenPRL2026}, demonstrated by scaling of evaporation branching 
ratios of water molecules.
This result was observed concurrently with the observation of complete hydrogen scrambling.
The water study has recently been extended to ammonia clusters~\cite{HansenPrep2026}, which 
showed a similar scaling behavior.

An obvious question is how these results carry over to clusters composed of molecules 
that lack the typical hydrogen bonding.
The methane molecule qualifies for this with its prototypical covalent bonds and 
was chosen as the clustering material for the present work. 

The presence of the molecule in space adds to the interest. 
In the deuterated form it is an important diagnostic tool for the dynamics and 
inventory of interstellar space~\cite{GibbIcarus2003}, a question posed much 
more powerfully with the very recent observation of an extremely high deuterium ratio 
in cometary methane~\cite{roth2026enrichedmethanedhratio}.
Beyond basic scientific curiosity, the methane molecule warrants attention qua its role
as a very potent greenhouse gas.
With the possibility of release of enormous amounts of frozen methane from thawing
tundras, every aspect of the condensed-to-gas transition merits 
study~\cite{FroitzheimaPNAS2021}.

\section*{Experimental methods and results}

The procedure used closely follows the one described in~\cite{HansenPrep2026} and 
the cryogenic protocol in~\cite{HansenPRL2026}. 
The charge necessary for the end-stage mass spectrometry of the experiment was 
generated by electron ionization of a helium droplet, followed by pick-up of 
a hydrogen molecule and transfer of the charge to a hydrogen atom.
Methane molecules were then picked up by the ~$10^7$ helium atom 
droplets~\cite{MauracherPR2018} where they aggregated into clusters. 
For this process, the two isotopologues \ce{CH4} and \ce{CD4} were introduced separately 
into the pick-up chamber. 
After cluster formation in the droplets, the helium atoms, being the most loosely 
bound entities in the droplets, were stripped from the clusters by collisional heating. 
Refs.~\cite{HansenPRL2026,HansenPrep2026,TiefenthalerPCCP2020,TiefenthalerRSI2020} 
give more details of the procedure.

\begin{figure}[htp]
	\includegraphics[width=\columnwidth]{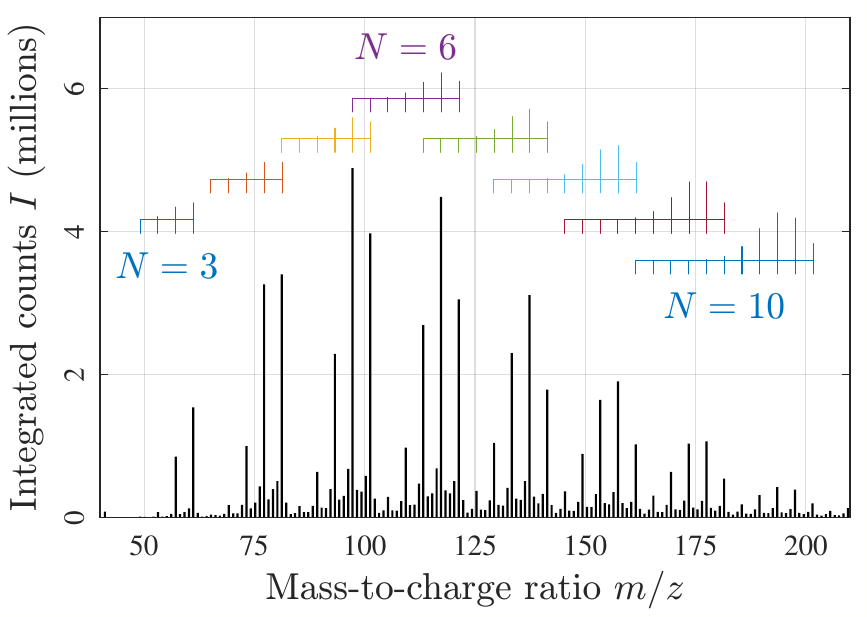}
	\caption{\label{fig:parent-mass-spectrum}%
		A measured mass spectrum of methane clusters produced with a pickup cell pressure 
		of $\SI{8.5e-5}{mbar}$. Binomial distributions are shown as colored combs at the 
		expected cluster masses, where $N$ is the number of methane molecules and the binomial 
		distribution parameter $p$ corresponds to a 1:3 mix of \ce{CH4} and \ce{CD4}. 
		The mass spectrum consists of strong peaks at the predicted masses, but also 
		low-intensity peaks corresponding to clusters with attached helium atoms.
		These are completely removed during the later collisional heating stage.}
\end{figure}

\begin{figure}[htp]
	\includegraphics[width=\columnwidth]{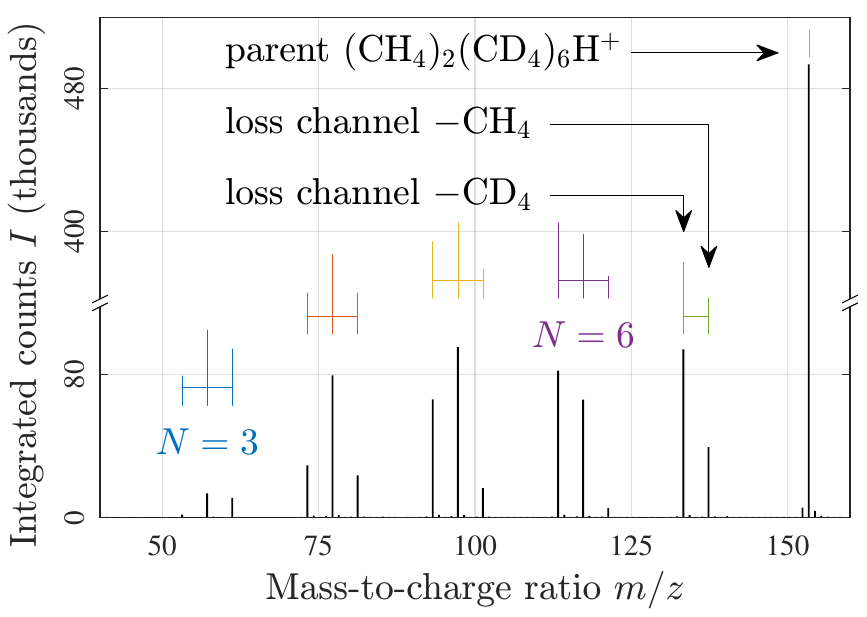}
	\caption{\label{fig:children-mass-spectrum} A mass spectrum of the cluster fragments 
	after collisional heating of an $N=8$ cluster. 
	Theoretical distributions assuming uniformly random evaporation are shown as colored 
	combs. 
	No other peaks are seen in the spectrum, implying that all helium has evaporated
	before the initial mass selection.
	Only single-molecule loss channels will be used in the analysis.}
\end{figure}

Figure~\ref{fig:parent-mass-spectrum} shows an example of a mass spectrum generated this way. 
The spectrum is processed from the raw time-of-flight mass spectrum data by integrating 
intensities over a mass interval of width 1 amu centered on each peak. 
Besides the clusters which show up with well separated isotopologue peaks, a small residual 
amount of helium attachments is also seen. 
This contamination was removed by the final collisional heating, as verified by the later 
mass spectra of the collision products.
After helium stripping, the clusters were mass selected and additional collisional heating 
was applied to induce evaporation of methane molecules.
After mass selection of one of the bare methane clusters produced in this last heating stage,
the time-of-flight mass spectrometer measured the lost mass.
Cluster sizes $N=8$--$10$ of different compositions were measured, i.e., 
\ce{(CH4)_$P$(CD4)_$D$H^+} with $P+D=N$. 
Figure~\ref{fig:children-mass-spectrum} shows an example of the products appearing 
after the collisional heating of an $N=8$ cluster. 
The single isotopologue molecules lost are indicated with black arrows. 

The isotope composition of the decay product allows us to quantify the degree of hydrogen 
scrambling in the cluster. 
None was observed. 
The isotopologue composition of the decaying clusters could then be determined exactly 
mass spectrometrically without the need for an intermediate step in the analysis to determine 
the populations arising from hydrogen scrambling, as was needed in~\cite{HansenPRL2026}.

\begin{figure}[htp]
	\includegraphics[width=\columnwidth]{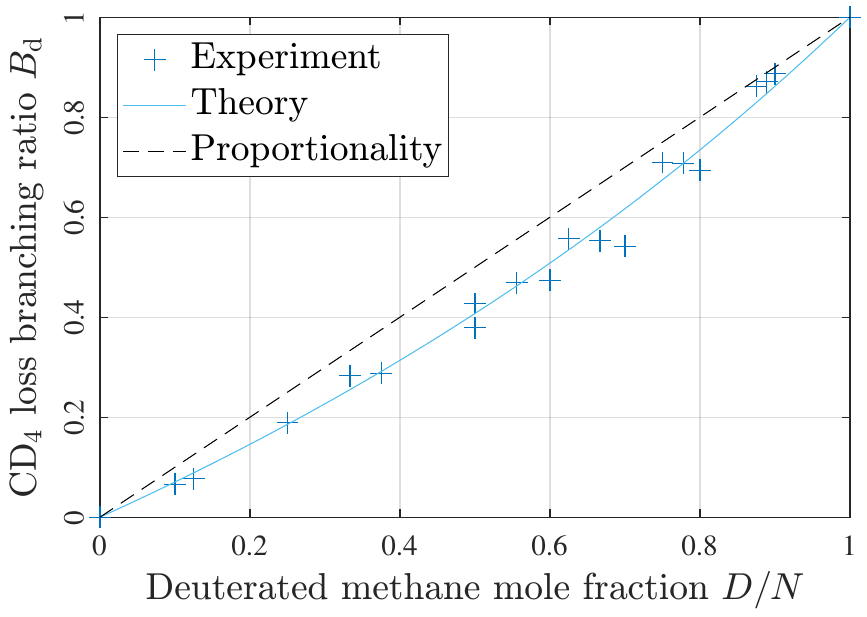}
	\caption{\label{fig:branching-ratio} %
		The branching ratio for deuterated methane loss, as a function of its mole 
		fraction in the parent. 
		The experimental values clearly deviate from the proportionality,
 	  and instead follow the full line, to be derived below.}
\end{figure}

The branching ratios for the light and heavy isotopologue loss are found from the 
integrated counts for the two channels, $I_{\rm p}$ and $I_{\rm d}$ for loss of protonated 
and deuterated molecules, respectively, as 
\begin{align}
	B_{\rm p} = \frac{I_{\rm p}}{I_{\rm p} + I_{\rm d}}
	\quad\mbox{and}\quad
	B_{\rm d} = \frac{I_{\rm d}}{I_{\rm p} + I_{\rm d}}.
\end{align}
The dependence of $B_{\rm d}$ on the composition of the parent species is shown in 
Fig.~\ref{fig:branching-ratio}. 
Clearly $B_{\rm d}$ is not simply proportional to its mole fraction in the parent. 
The deviation from the direct proportionality nevertheless scales, however, 
with the data for 
all three parent cluster sizes measured falling on the same curve, a behavior similar 
to that of water and ammonia clusters~\cite{HansenPRL2026,HansenPrep2026}, 
although appearing with the admittedly lower statistics here.

\section*{Data analysis}

\begin{figure}[htp]
	\includegraphics[width=\columnwidth]{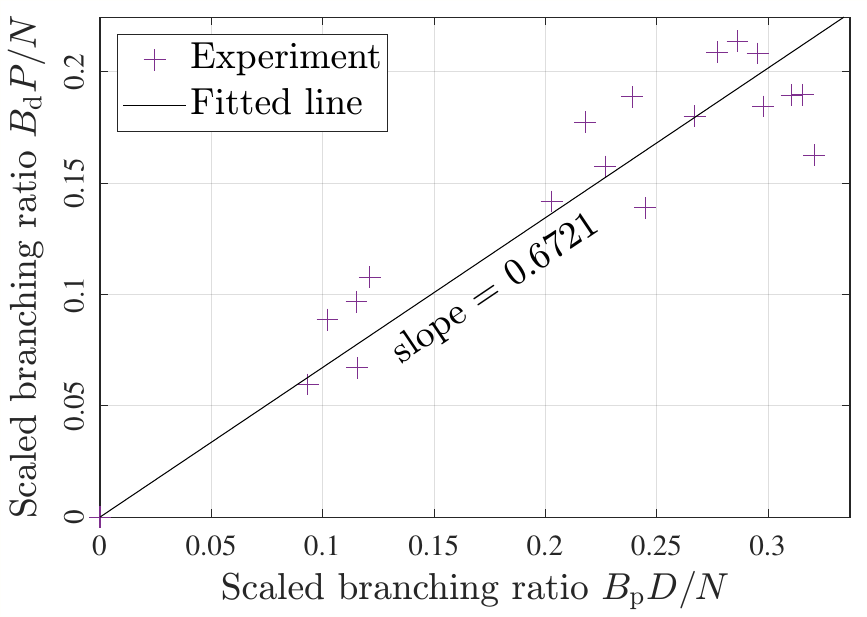}
	\caption{\label{fig:fitted-line}%
		The two branching ratios plotted against each other, after multiplication with 
		the population of the mole fraction of the other species. 
		The fitted line is the linear regression constrained to passing through the origin.}
\end{figure}

The scaling seen in the branching ratio represents rate constants multiplying 
the respective mole fractions, $P/N$ and $D/N$ for undeuterated and deuterated 
molecules:
\begin{align}
	B_{\rm p}\quantity(D/N) &\propto k_{\rm p}\quantity(D/N)\frac{P}{N}\\
	B_{\rm d}\quantity(D/N) &\propto k_{\rm d}\quantity(D/N)\frac{D}{N},
\end{align}
where $P/N$ and $D/N$ are the measured isotopologue mole fractions and 
$k_{\rm p},k_{\rm d}$ are the light and heavy methane evaporation rate constants, 
respectively.
The scaling seen in Fig.~\ref{fig:branching-ratio} is consistent with a constant 
ratio of the two $k's$.
This is verified by plotting the quantities $B_{\rm d}P/N$ and $B_{\rm p}D/N$ vs. 
each other (this representation of the data is chosen to avoid division by near-zero 
mole fractions).
Figure~\ref{fig:fitted-line} shows the plot which yields the value $1.47 \pm 0.08$ for 
$k_{\rm p}/k_{\rm d}$.

The proportionality in Fig.~\ref{fig:fitted-line} only demonstrates a constant ratio 
of the $k$'s, and does not exclude a common variation with $D/N$. 
Such a variation will in fact arise because the effective emission temperature 
varies with the mixing ratio. 
This variation arises due to the method of preparation of the clusters.
Consider the process of raising the temperature from the 0.4~K helium 
droplet temperature with the helium stripping and the subsequent collisional 
cluster heating process.
The energy acquired during these processes develops a width due to both varying 
numbers of collisions and different amounts of energy transfer in each collision.
Evaporation occurs at or close to the excitation energy that fulfills the relation
$k(E) = 1/t$~\cite{hansen01,HansenMSR2021} 
where $t$ is on the order of the time elapsed between last collisional heating and 
the measurement in the ToF-MS.
The final numbers are not very sensitive to the precise time and we will simply use 
the value $t=50~\mu$s.
With the effective decay constant for a cluster of composition $P,D$ of
$k(E) = \frac{P}{N}k_{\rm p}(E) + \frac{D}{N}k_{\rm d}(E)$ the composition-dependent
energy of the decaying clusters is picked out by the relation
\begin{align}
	\label{T-det}
	\frac{P}{N}k_{\rm p}(E) + \frac{D}{N}k_{\rm d}(E) = 1/t.
\end{align}
In general $k_{\rm p}$ and $k_{\rm d}$ have different dependencies on energy.
Hence the solution for $E$ will depend on the precise value of $D/N$.

Detailed balance rate constants were essential to understand similar data on ammonia
clusters and we will use such rate constants here also to calculate the frequency factor.
With an Arrhenius form we have, following the procedure in ref.~\cite{HansenPrep2026}:
\begin{align}
	\label{rateconst}
	k_i&=\frac{\mu_i}{\pi^2 \hbar^3} \sigma Z_{{\rm rot},i} 
	k_{\rm B}^2 T_{N-1}^2 \\\nonumber&\times
	\exp\quantity( -\frac{E_{{\rm a},i}}{k_{\rm B}T(E-E_{{\rm a},i}/2)}),
\end{align}
where $i$ is p or d, $\mu_i$ is the reduced mass of the channel, $\sigma$ the 
attachment cross section for the inverse process, and $Z_{{\rm rot},i}$ the 
rotational partition function of the evaporated molecule.
$T_{N-1}$ is the product cluster temperature, defined as the reciprocal of the 
logarithmic derivative of the product level density, with the reactant temperature 
in the exponential defined similarly.
The effective temperature is proportional to $E_{{\rm a},d}$ and at the price of an 
added interpretation of the temperatures, which are not measured experimentally,
we can replace both $T_{N-1}$ and $T(E-E_{{\rm a},d}/2)$ by a single 
temperature, $T$, absorbing the finite heat bath correction~\cite{KlotsJPC1988} of 
the reactant to give
\begin{align}
	\label{rateconst2}
	k_i=\frac{\mu_i}{\pi^2 \hbar^3} \sigma Z_{{\rm rot},i} 
	k_{\rm B}^2 T^2 
	\exp\quantity( -\frac{E_{{\rm a},i}}{k_{\rm B}T}).
\end{align}
Further, for the reduced mass we can use the relevant molecular mass, 16 amu or 20 amu.
For the cross section we use $\sigma = \pi r_1^2 (8^{1/3}+1)^2 = 3.8\times 10^{-19}$~m$^2$ 
corresponding to the capture of a monomer by the octamer with a geometric cross section,
and $r_1$ is the monomer radius determined from the bulk molecular number density 
$\nu$ as $1/\nu = 4\pi r_1^3/3\times 16~{\rm amu}$, with 
$\nu = 422.4$~kg/m$^3$~\cite{HandbookCP}.
For the $T$ in the pre-exponential the bulk enthalpy of vaporization at the atmospheric
vapor pressure, 510.8~kJ/kg, converted to the per-molecule value 
$k_{\rm B} \times 983$~K = 84.7~meV and divided by $\ln(\omega_i t)$, is used. 

Finally, the rotational partition functions depend on the quantum statistics of 
nuclear spins and spatial rotations.
The simple recipe of classical values divided by the symmetry numbers which holds 
in the high-temperature limit~\cite{ludloff1929gleichgewicht} is marginally accurate 
at the temperatures relevant here.
We instead used interpolations between the numerically calculated values in 
ref.~\cite{Robiete1979}.
This determines the parameters of the pre-exponential and gives the two rate constants
\begin{align}\label{kp}
	k_{\rm p} = 5.50\times 10^{14}~{\rm s}^{-1} \exp\quantity(-E_{\rm a,p}/T ),\\
	k_{\rm d} = 1.61\times 10^{15}~{\rm s}^{-1} \exp\quantity(-E_{\rm a,d}/T ),\label{kd}
\end{align}
for a temperature at or close to 40~K.
A few algebraic manipulations give the ratio of evaporative activation energies
\begin{align}
	\label{E-ratio}
	\frac{E_{{\rm a,d}}}{E_{{\rm a,p}}} = 1 + 
	\frac{
	\ln\quantity(\frac{\omega_{\rm d} k_{\rm p}}{\omega_{\rm p} k_{\rm d}})}{\ln{\omega_{\rm p}t}} 
	= 1.0607.
\end{align}

\begin{figure}[htp]
	\includegraphics[width=\columnwidth]{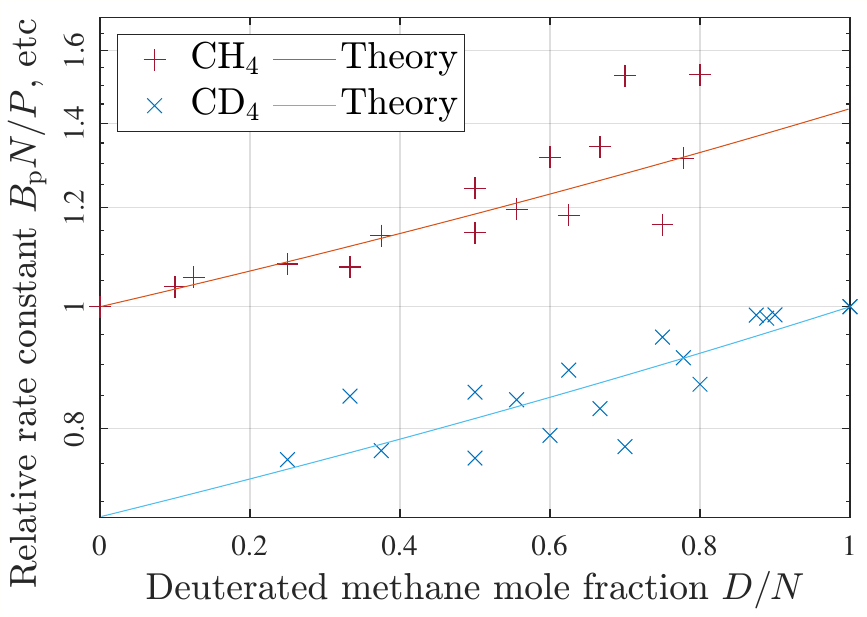}
	\caption{\label{fig:rate-constant}The relative values of the two rate constants for 
	light (red horizontal crosses) and heavy (blue tilted crosses) isotopologue 
	emission vs cluster composition.
	The two lines, calculated in the main text, differ by a factor of 1.42, close 
	to the expected 1.47.
	The experimental points are limited to mole fractions above 0.2 to eliminate 
	statistical fluctuations.}
\end{figure}

We are now ready to answer the question of the mixing ratio dependence of a common 
factor for the two $k$'s.
Figure~\ref{fig:rate-constant} shows the two experimental branching ratios divided 
by their respective mole fractions.
The separation between protium and deuterium points is consistent with the fitted factor 
of 1.47 of Fig.~\ref{fig:fitted-line} across the range of $D/N$ from zero to unity, 
indicating that a shared $D/N$-dependent factor indeed is the case and justifies 
the treatment of the $k$'s as relative constants in the specific plot in 
Fig.~\ref{fig:fitted-line}. 

The multiplicative dependence of $D/N$ can be found by application of Eq.~\ref{T-det} 
with the expressions for the $k$'s from Eqs.~\ref{kp},~\ref{kd}.
The temperature is found numerically and provides the temperature for the two rate 
constants as a function of $D/N$. 
These are shown in Fig.~\ref{fig:rate-constant} as the two thin lines through the 
corresponding experimental points.
It should be stressed that the calculated curves rely only on the calculated 
ratio of the frequency factors and the ratio of the two rate constants fitted from
the experimental data in Fig.~\ref{fig:fitted-line}.
In particular, no adjustable parameter was used to produce the curves.
The agreement should be judged on the fact that a change of the ratio 1.0607 
by just one promille will give an observably worse agreement.
The same parameters are also used to calculate the branching ratio curve in 
Fig.~\ref{fig:branching-ratio}.
The agreement between measured and calculated curves is again very satisfactory.

\begin{figure}[htp]
	\includegraphics[width=\columnwidth]{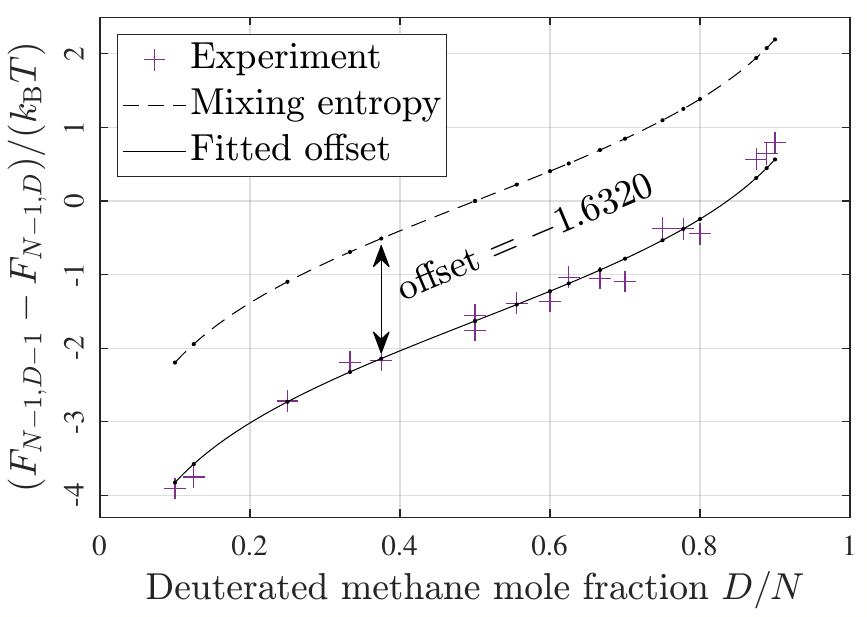}
	\caption{\label{fig:free-energy-diff} The free energy differences of 
	(CH$_4$)$_{P-1}$ (CD$_4$)$_{D}$ and (CH$_4$)$_{P}$ (CD$_4$)$_{D-1}$ 
	vs. the deuterated mole fraction of the parent (crosses)
	and the mixing entropy of the composition of the parent (dashed line).
	Small dots on the full line give the parent mixing entropy minus $1.63$.}
\end{figure}

Finally, the difference in the Helmholtz free energies of the products of the 
evaporation of the two isotopologue molecules can be calculated, as outlined 
in~\cite{HansenPRL2026} and derived in more detail in~\cite{HansenPrep2026}.
The exponential of the difference in free energies is given as
\begin{align}\label{FreeEnergies}
	\exp\quantity(-\beta (F_{\rm d} - F_{\rm p})) = 
	\frac{B_{\rm d}/\omega_{\rm d}}{B_{\rm p}/\omega_{\rm p}},
\end{align}
where the difference in the free energy here is 
\be
F_{\rm d} - F_{\rm p} = E_{\rm a,p} - E_{\rm a,d} -T\(S_{\rm p} + S_{\rm d}\),
\ee
and hence differs from the macroscopic expression by the absence of the thermal energy
above the respective ground state energies.
Figure~\ref{fig:free-energy-diff} shows the right hand side of Eq.~\ref{FreeEnergies} 
vs. the deuterated mole fraction together with the calculated differences in mixing 
entropies, $\Delta S$.
This difference involves simple combinatorics, resulting in the form
\begin{align}
	\exp\quantity(\Delta S ) = \frac{D/N}{1-D/N}.
\end{align}
Figure~\ref{fig:free-energy-diff} also shows this calculated mixing entropy shifted down by 1.63.
This aligns the calculated curve with the measured points very well 
and gives a value for the separation energy difference of 
$\Delta E = k_{\rm B}T \times 1.63$ for the two channels. 
With the temperature from Eq.~\ref{T-det} (at $D=N$) of 
$0.042 E_{\rm a,p}/k_{\rm B}$ we get
\begin{align}
	\Delta E = 0.068 E_{\rm a,p}.
\end{align}
The value compares well with the value found in Eq.~\ref{E-ratio}.
\begin{align}
	\Delta E = 0.061 E_{\rm a,p}.
\end{align}

\section*{DFT Calculations}
For the purpose of obtaining absolute values of dissociation energies, 
density functional theory (DFT) calculations were performed. 
The theoretical pipeline consisted of three stages: 1) generation 
of structures, 2) filtering based on energy, and 3) final evaluation including 
vibrational zero-point energies (VZPE).

In the first stage, a basin-hopping algorithm sampled the space of possible 
clusters by repeatedly mutating the unique lowest-energy structures of each 
cluster size. 
The mutations consisted of deletion or addition of monomers, followed by the 
PM7 method as implemented in MOPAC~\cite{moussa2026mopac}. 
The terminating condition was that, for every cluster size $N$ between 3 and 
10, the $N^2$ unique lowest-energy structures had been subjected to $N$ additions 
and $N$ distinct deletions, all distinct. 
Upon termination, $\SI{54000}{}$ structures had been found, of which 
$\SI{34000}{}$ were of experimentally relevant cluster size $N=8$--$10$.

In the second stage, the 1000 lowest-energy structures ($\approx\SI{3}{kcal/mol}$ 
window) of each cluster size were loosely optimized with Gaussian~\cite{frisch2016g16}, 
using the B3LYP functional including GD3BJ empirical dispersion over the cc-pVDZ basis set. 
The remaining structures were discarded, implicitly assuming that PM7 and B3LYP energies
were sufficiently correlated. 

In the third and final stage, the 100 lowest-energy structures were tightly optimized 
using the same functional over the Aug-cc-pVTZ basis set. 
After optimization, the Hessian (force constants) matrix was calculated and 
subsequently combined with mass arrays to yield the zero-point energies for 
all $2^N$ isotopologues of the ground state.
The vibrational frequencies were scaled with empirical dual factors fitted to monomer 
data~\cite{UlenikovJCP2014}.  
They were 0.9813 and 0.9645, below and above 2000 cm$^{-1}$, respectively.

\begin{figure}[htp]
	\centering
	\includegraphics[width=\linewidth]{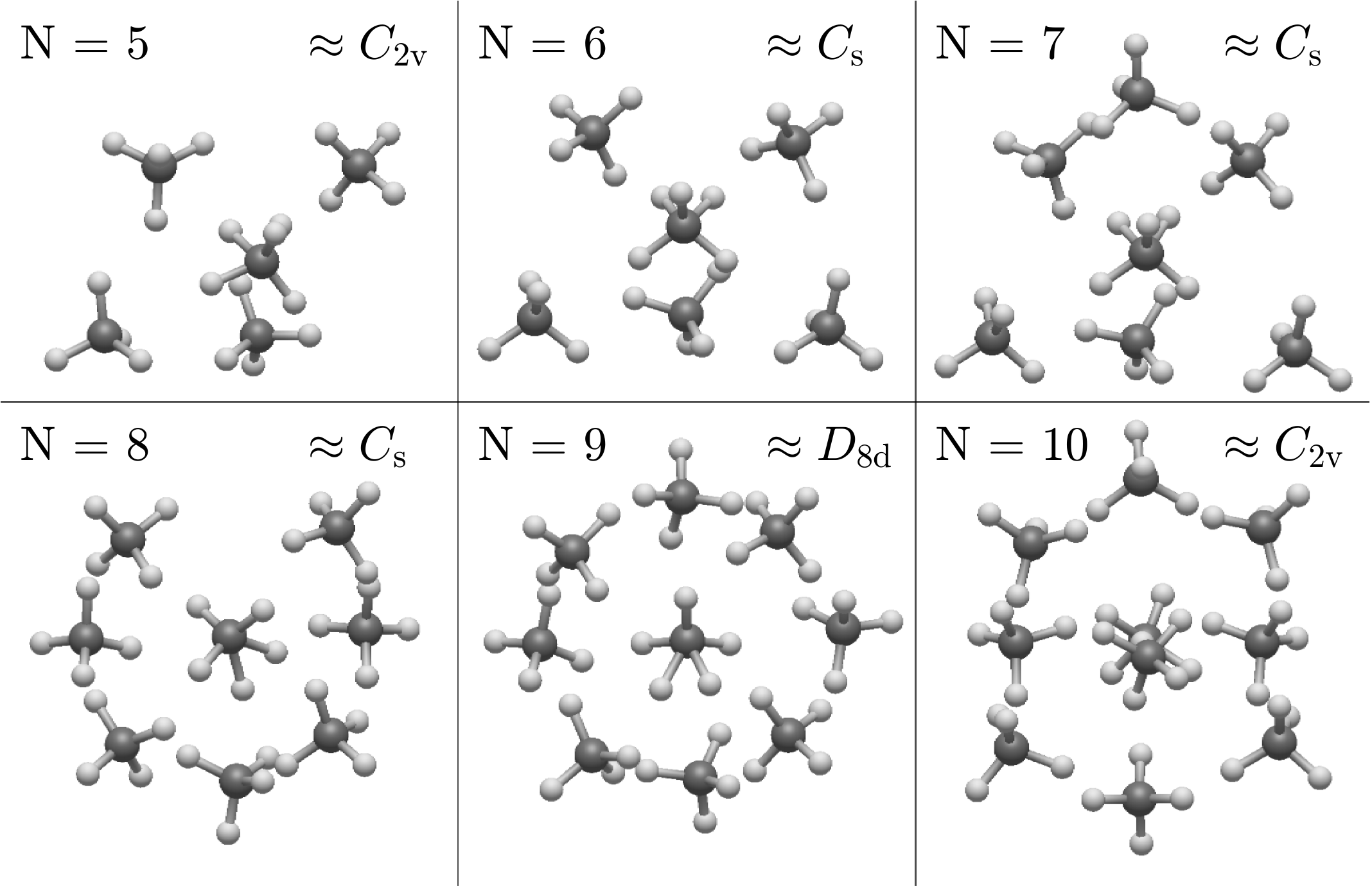}
	\caption{\label{fig:methane-clusters}%
		The ground state protonated clusters of methane, sizes $N=5$--$10$. 
		The surplus proton is always bound to the central methane. 
		Although no cluster shown has a strict symmetry, their carbon subsets have the 
		approximate symmetries indicated in the frames.
	}
\end{figure}

Figure~\ref{fig:methane-clusters} shows the found ground states of the protonated clusters. 
The structures are stabilized by the protonated methane, which tends to sit in the center 
and maximize its coordination number. 
The nonamer appears to have a closed shell of 8 
equivalent neutral methane in a $D_\mathrm{8d}$-symmetric square antiprism pattern.

\section{DFT Data Analysis}
The desired dissociation energies depend on the specific decaying isotopologue cluster. 
In order to get a single effective number, we fit a model to the computed energies and 
calculate the dissociation energies analytically.

The simplest model for energy is
\begin{align}
	E_\mathrm{b} = \alpha_1 + \beta_1 N + \gamma_1 D.
\end{align}%
\begin{figure}[htp]
	\centering
	\includegraphics[width=\linewidth]{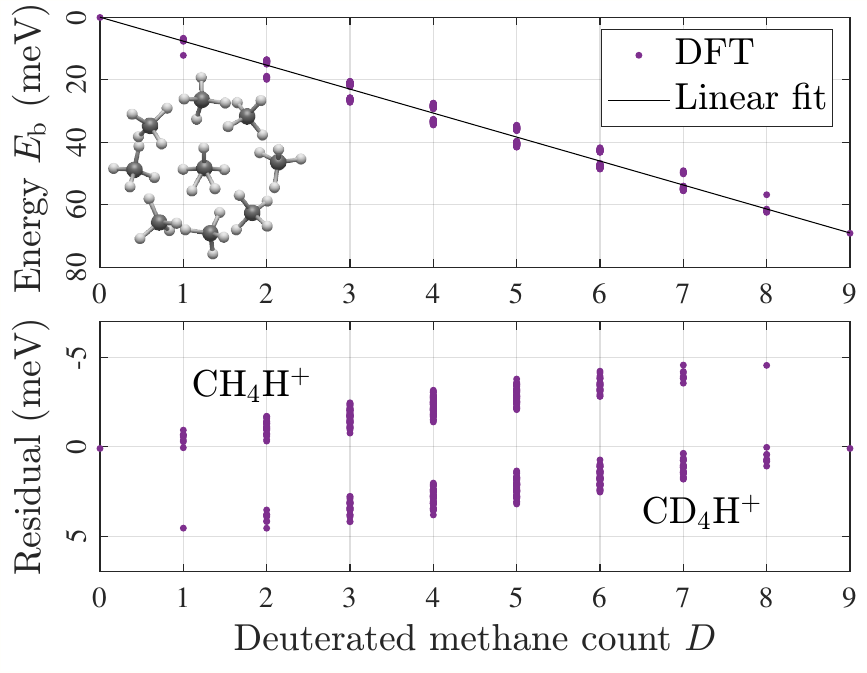}
	\caption{\label{fig:dft-energy-nonamer}%
		Top: computed binding energies $E_\mathrm{b}$ of the nonamer isotopologues, 
		relative to the all-protium ($D=0$) nonamer. 
		The binding energy is defined as the cluster energy subtracted from the sum 
		of monomer energies. 
		A linear fit $E_\mathrm{b}(D)$ is shown in black. 
		Bottom: the residuals show a strong dependence on where the surplus proton 
		sits, \ce{CH4} or \ce{CD4}.}
\end{figure}%
Figure \ref{fig:dft-energy-nonamer} shows this model fitted to the isotopologues of 
the nonamer. 
While the $R^2$ measure is quite good, the residuals show a strong dependence on 
whether the surplus proton sits on a light or heavy methane. 
We therefore introduce the number $D_5$, which describes whether the surplus proton 
sits on a light ($D_5=0$) or heavy ($D_5=1$) methane, and note that the number of 
deuterated methane molecules without the surplus proton is $D_4 = D-D_5$.
The cluster formula is then
\begin{align}
	\ce{(CH4)_$N-1-D_4$(CD4)_$D_4$(CH4H^+)_$1-D_5$(CD4H^+)_$D_5$},
\end{align}
where the notation $\ce{CH4H^+}$ emphasizes that the surplus proton is distinguishable 
from the four covalently bound hydrogen atoms~\cite{sefcik1974methanium}.

The next simplest model for energy is then
\begin{align}
	E_\mathrm{b} = \alpha_2 + \beta_2 N + \gamma_2 D_4 + \delta_2 D_5.\label{eq:dftfit2}
\end{align}%
\begin{figure}[htp]
	\centering
	\includegraphics[width=\linewidth]{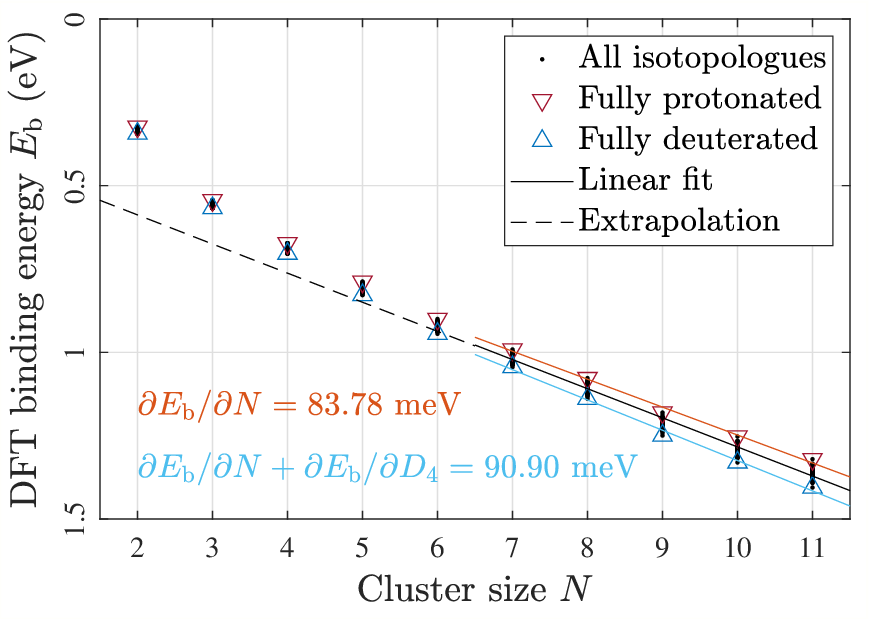}
	\caption{\label{fig:dft-energy-size}%
		Computed binding energies $E_\mathrm{b}$ of all isotopologues, together with 
		the fit from Eq.~\eqref{eq:dftfit2}. involves      
		The fully protonated (dark orange) and deuterated (blue) clusters are highlighted 
		with circles. 
		The linear model is fitted to clusters with $N=7$--$11$ and is shown as three 
		lines, corresponding to the restrictions $(D_4,D_5)=(0,0)$ (light orange), 
		$(D_4,D_5)=(N-1,1)$ (blue), and $(D_4,D_5)=(N/2-1/2,1/2)$ (black). 
		The slopes of the outer lines are written out and correspond to the dissociation 
		energies $E_\mathrm{a,p}$ (orange) and $E_\mathrm{a,d}$ (blue).}
\end{figure}%
Figure~\ref{fig:dft-energy-size} shows this model fitted to all isotopologues of 
cluster sizes $N=7$--$11$. 
Smaller clusters are omitted because of their atypically low coordination numbers 
and thus binding energies. 
The residuals (not shown) are not systematic with respect to $D_4$ or $D_5$, but show 
some dependence on $N$. 
In particular, the symmetric nonamer has a binding energy $\SI{16}{meV}$ stronger than 
the linear model would predict.

The main theoretical result is the dissociation energies, which are obtained by 
differentiating the binding energy
\begin{align}
	E_\mathrm{a,p} &= \partial E_\mathrm{b} / \partial N &&\hspace{-2em}= \SI{83.78}{meV} \\
	E_\mathrm{a,d} &= \partial E_\mathrm{b} / \partial N 
	+ \partial E_\mathrm{b} / \partial D_4 &&\hspace{-2em}= \SI{90.90}{meV}.
\end{align}
The calculated CH$_4$ dissociation energy $E_{\rm a,p}$ is remarkably close to the 
experimental bulk value of 84.7 meV~\cite{HandbookCP}.
The calculated ratio $E_\mathrm{a,d}/E_\mathrm{a,p} = 1.085$ is slightly higher 
than the experimental values of $1.061$ and $1.068$, but still by a fairly small 
number considering the large energies involved in the numerical derivative.
This ratio would have been 1.087 without empirical scaling factors.
In absolute numbers the difference between the experimental and theoretical results 
of the ratios 1.085 and 1.0645 amounts to 1.7~meV.

\section{Discussion}

There are several sources for the difference between protonated and deuterated methane 
loss rates.
The most important difference between the two isotopologues is their mass,
and the only one that is considered in the quantum chemical calculations.
The masses have obvious consequences for the vibrational frequencies, both 
intra- and intermolecular types.
These differences will translate into differences in binding energy directly via the
differences in zero-point energies of cluster-bound and free molecules.
The total difference between the CH$_4$ and CD$_4$ evaporative activation energies,
calculated to be 7.12~meV, is similar to the calculated difference of 
vibrational zero point energies (VZPE) of 7.86~meV for the same decays of the nonamer.

The vibrational spectra of the clusters separate into two groups of the inter-molecular 
at low frequencies and the intra-molecular frequencies, the latter appearing
in two groups centered at frequencies around the molecular values of the isotopologues.
The dominant contribution to the difference of VZPE for protonated and deuterated molecules
is provided by the intermolecular vibrations.
The lowest (intermolecular) 42 modes of the two octamer species discussed here differ 
by 7.83~meV, which can be compared directly with the overall 7.12~meV activation energy 
difference. 
It is practically identical to the total VZPE difference of 7.86~meV, indicating that 
cluster effects on the average intra-molecular vibrational frequencies is minute,
and that total VZPE differences and activation energy differences are given by the 
intermolecular differences alone.

Finally, the magnitude of the intra-molecular modes is influenced by the possibility of 
free molecular rotations.
Methane shows free molecular rotations also at low temperatures~\cite{SlipchenkoJPCL2016,
HoshinaJCP2015,ProkhorovJPCS2019}, at least for molecules located at the cluster surface,
and even at the 0.4 K helium droplet molecular beam temperature.
This effect is not included in the diagonalization of the Hessian used to find the VZPE 
but also appears not to have a significant consequences for the numbers for the measured 
clusters.

\section{Summary}

We have presented experimental results on the distillation of methane for two heavy and light 
methane from clusters. 
The results show firstly that there is no exchange of hydrogen isotopes of proton and deuterium.
The evaporative branching ratios can be represented by a simple ratio of light and heavy 
methane evaporation rate, multiplied by the mole fraction in the decaying cluster. 
Experimental branching ratios are reproduced without any need for a fitting parameter, and 
relative activation energies for the two processes are determined be $1.07\pm 0.01$.
Finally, the difference in mixing entropy of the two clusters produced in the parallel channel 
evaporation processes is measured.
The results for all three clusters measured fall on the same curve without application of 
any adjustable parameter.

%

\section*{acknowledgments}
%

\end{document}